\documentclass[a4paper]{article}

\usepackage[pages=all, color=black, position={current page.south}, placement=bottom, scale=1, opacity=1, vshift=5mm]{background}
\SetBgContents{
}      

\usepackage[margin=1in]{geometry} 

\usepackage{amsmath}
\usepackage{amsthm}
\usepackage{amssymb}

\usepackage{bbm} 

\usepackage[utf8]{inputenc}
\usepackage[colorlinks,citecolor=blue,urlcolor=blue,filecolor=blue,backref=page]{hyperref}
\hypersetup{
	unicode,
	pdfauthor={Author One, Author Two, Author Three},
	pdftitle={A simple article template},
	pdfsubject={A simple article template},
	pdfkeywords={article, template, simple},
	pdfproducer={LaTeX},
	pdfcreator={pdflatex}
}

\usepackage{amsthm}
\usepackage{amsmath}
\usepackage{natbib}
\usepackage{graphicx}

\title{Sample size reassessment in Bayesian hybrid clinical trials}
\author{Marco Ratta$^{1,2}$, Pavel Mozgunov$^{1,3}$, Sandrine Boulet$^{4,*}$, Moreno Ursino$^{4,*,**}$}

\date{
	$^1$Department of mathematical science (DISMA), Polytechnic University of Turin, Turin, Italy \\%
	$^2$Department of Statistical Methodology, Saryga, Tournus, France \\%
	$^3$MRC Biostatistics Unit, University of Cambridge, Cambridge, UK\\%
	$^4$ Inserm, UMRS 1346, Inria, Université Paris Cité, F-75015 Paris, France\\%
	$^{*}$Sandrine Boulet and Moreno Ursino made equal contributions and are co-last authors.\\%
	$^{**}$Corresponding author.\\%
}

\begin{document}
\maketitle

\begin{abstract} 
The use of historical controls offers a valuable alternative when traditional randomized controlled trials are not feasible. However, such approaches may introduce bias due to temporal changes in patient populations, diagnostic criteria, and/or treatment standards. Hybrid designs, which combine a concurrent control arm with historical control data, can help mitigate the possible bias.
We propose a novel Bayesian two-arm randomized clinical trial design incorporating an interim analysis. At the interim analysis, a new criterion derived from the Hellinger distance is used to quantify the similarity between historical and concurrent control data outcomes. This measure informs both (1) the variance function of the control prior distribution in the final analysis and (2) the sample size reassessment for the second stage of the trial.
The proposed approach is designed to accommodate both continuous and binary endpoints and is assessed through extensive simulation studies. Results demonstrate the method flexibility and robustness in adapting to varying degrees of historical-control heterogeneity.

\end{abstract}

\section{Introduction}\label{sec1}

Randomized controlled trials (RCTs) are the gold standard for evaluating the efficacy and safety of new medical interventions~\citep{lee_2005, wieand_2005, schulz_2010, hariton_2018}. Randomization, when the sample size is sufficient, ensures that known and unknown prognostic factors and effect modifiers are balanced between treatment and control arms, and thus limits selection bias. However, in certain situations, conducting a traditional RCT may be challenging or unethical. Indeed, RCTs require a large sample size, making the study longer and more costly. Furthermore, a large sample size is almost an impossible requirement for clinical trials involving rare diseases. As a result, a lot of Phase II trials are built using only a single-arm, that is without control arm. In such cases, the use of historical controls (HCs) can provide a valuable alternative approach. HCs are data derived from patients who received a standard treatment (or no treatment in case of placebo controlled trials) in the past, and their outcomes are compared with those of patients receiving the new intervention in the current study~\citep{viele_2014, marion_2023}. This approach can be particularly useful when it may be difficult to recruit a sufficient number of patients for a traditional RCT, or in situations where a placebo control group is not ethical. The use of HCs in clinical trials has a long history, with early examples dating back to the 1940s. The methodology for using HCs has evolved significantly over time, along with the increasing recognition of the potential biases that can arise from comparing data from different time periods. These biases can include changes in patient populations, diagnostic criteria, and treatment standards over time. To address these challenges, various statistical methods, including the use of 
propensity scores related methods~\citep{ross_2015} or counterfactual analysis approaches~\citep{rippin_2022}, have been developed to adjust for potential confounding factors and improve the reliability of comparisons between HCs and current study participants. 
However, a recent cross-sectional study~\citep{liu_2025} investigating current practices in the design, conduct, and analysis of externally controlled trials shows that these methods are still not widely used. 
Despite the potential limitations, HCs can offer several advantages in specific clinical trial settings. In particular, they can \textit{i)} reduce the sample size required for a study, potentially leading to faster and more efficient evaluation of new treatments and \textit{ii)} help to minimize the number of patients exposed to a placebo or an ineffective treatment. Indeed, in July 2025, the European Medicines Agency launched a public consultation to write a discussion paper describing the principal challenges associated with external controls and to examine in more detail the circumstances and methodological constraints for which using external controls could be considered appropriate for generating pivotal or supportive evidence, whether in terms of efficacy, safety, or other objectives relevant to regulatory decision-making. 

Relying exclusively on HCs may be statistically problematic, as this could introduce bias. 
A natural way to mitigate this issue is through the use of a hybrid design~\citep{ventz_2022}, which integrates historical control data with the concurrent control arm of the RCT. By supplementing the control group with all or part of historical control groups, more RCT participants can be allocated to the treatment arm compared to a traditional RCT. However, it is essential to ensure that the historical and current control data are sufficiently comparable to maintain the validity of the results.
For example, \cite{jiang_2023} and \cite{guo_2024} work in a causal framework, using a borrowing method based on propensity score, and require historical control patient-level data available.

The Bayesian framework naturally fits the use of HC in a statistical analysis. Indeed, Bayesian prior distributions can be used to incorporate external control data into a clinical trial. There are a number of different methods for building Bayesian prior distributions from external control data. For example, the Power Prior method~\citep{ibrahim_2015} raises the likelihood of the historical data to a power parameter ranging from 0 to 1, which controls the weight given to past information. The choice of the power parameter allows for borrowing strength while mitigating potential biases from differences between past and present data~\citep{gravestock_2017, ollier_2020}. Another common class of prior distributions is the commensurate prior, which introducing a commensurability parameter~\citep{hobbs_2011}. The Meta-Analytic Prior (MAP)~\citep{schmidli_2014} synthesizes prior information from multiple sources, typically through hierarchical modeling, capturing both the central tendency and variability across studies. To address potential conflicts between the prior and the data, the robust MAP is constructed as a mixture of the MAP and a non-informative distribution.
These methods have been used mostly to decrease the number of patient recruited in the control or placebo arm~\citep{hueber_2012, glatt_2018,baeten_2013}. 
While the latter approaches rely on parametric mechanisms to regulate borrowing, non-parametric alternatives have also been proposed, such as the covariate-balancing Bayesian-bootstrap approach of~\cite{wang_tiwari_2025}.

\begin{figure}[tb]
\begin{center}
  \includegraphics[width=0.3\textwidth,  angle=-90]{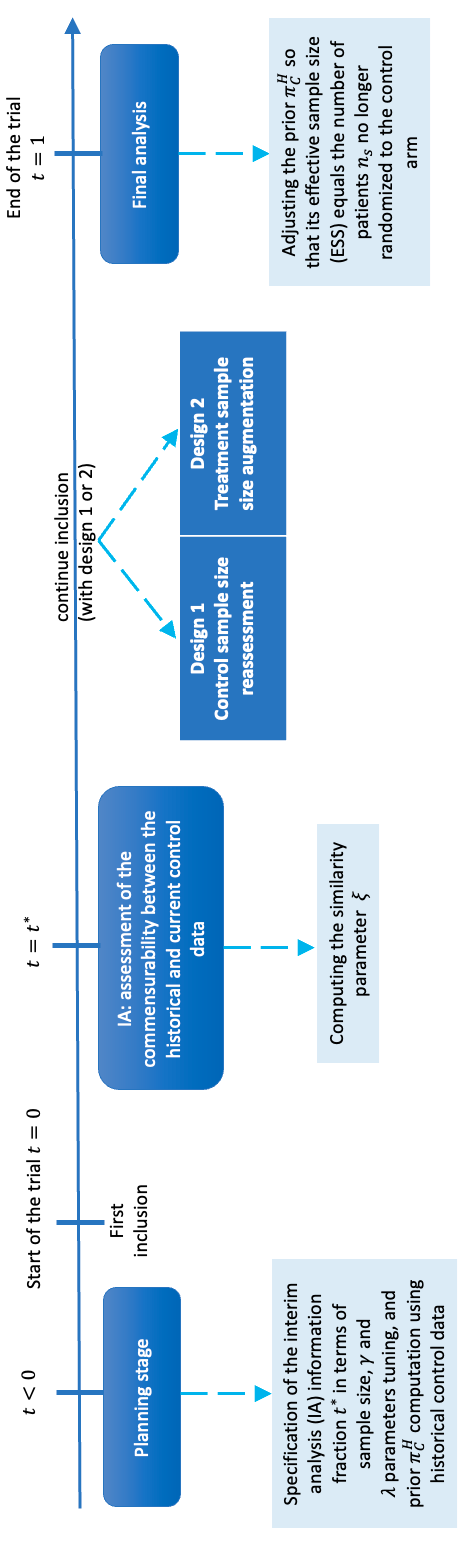}
  \caption{Schema of the proposed approach. A two-arm randomized clinical trial design is considered, including an interim analysis (IA) to assess the commensurability between the historical control and the current control data, potentially adjusting the randomization scheme after the IA. In this context, a new metric derived from the Hellinger distance is proposed to measure the degree of dissimilarity at the time of the IA. Then, this metric is used both (1) as a parameter of the variance function of the control prior distribution for final analysis and (2) to reassess the sample sizes for the second stage.}\label{fig:methods}
\end{center}
\end{figure}

In this manuscript, we consider a two-arm randomized clinical trial design. However, following \cite{ventz_2022}, we include an interim analysis (IA) to assess the commensurability between the historical control and the current control data, potentially adjusting the randomization scheme after the IA. In this context, a new criterion derived from the Hellinger distance is proposed to measure the degree of dissimilarity at the time of the IA. Then, this metric is used both (1) as a parameter of the variance function of the control prior distribution for the final analysis and (2) to reassess the sample sizes for the second stage (see Figure \ref{fig:methods}). 
Note that the use of the Hellinger distance in hybrid trial designs is not novel; for example, ~\cite{lu_2024} recently employed the Bhattacharyya coefficient (which is a metric deterministically related to the Hellinger distance) to define the borrowing parameter in a power-prior framework. However, its use in the context of adaptive trials with sample size reassessment has not been explored before.

This paper is organized as follows. The methods are given in Section \ref{sec2}. Section \ref{sec3} describes the simulation settings used to assess our methodology and the simulation results. A practical application based on a case study is described in Section \ref{sec4}. A conclusion closes the manuscript in Section \ref{sec5}.

\section{Methods}\label{sec2}

Consider a RCT where a treatment (T) is compared to a placebo or standard of care (C), with \( N \) patients allocated using an \( R:1 \) randomization ratio (T vs. C). Suppose the patient's treatment response \( Y_T \) and control response \( Y_C \) follow some underlying probability distributions with unknown parameters. For example, continuous variables \( Y_k \), for \( k \in \{T, C\} \), may follow a normal distribution, \( Y_k \sim N(\mu_k, \sigma_k^2) \), while dichotomous variables may follow a Bernoulli distribution, \( Y_k \sim \mbox{Bernoulli}(p_k) \). Without loss of generality, we assume that if \( Y_k \) is a continuous variable, it has been normalized so that \( \sigma_k = 1 \). In this work, the treatment effect is defined as the difference \( \Delta \) in expected mean responses between the treatment and control groups, $\mu_T - \mu_C$ with \( \mu_k = E[Y_k]\). The objective of the trial is to test the null hypothesis \( H_0: \Delta = 0 \) against the alternative \( H_A: \Delta > 0 \), indicating a positive treatment effect.  

In a Bayesian framework, prior distributions must be specified for the unknown parameters underlying the treatment and control responses. Typically, limited prior knowledge is available for the treatment parameter, in which case a weakly informative distribution is often chosen. For continuous outcomes, a normal prior distribution \( \pi^0_T \) can be assigned to \( \mu_T \), with a mean \( m^0_T \) and a standard deviation \( \sigma^0_T = 1 \) (also known as the unit information prior in this case with \( \sigma_k = 1 \)). For a binary variable, a Beta distribution parameterized by its mean \( m^0_T \) and precision parameter \( \phi^0_T \) can be assigned to \( p_T \), i.e., \( p_T \sim \mbox{Beta}(m^0_T, \phi^0_T) \), where \( \phi^0_T \) is typically set to 1 or 2. Note that \( \sigma \) and \( \phi \) are directly linked to the Effective Sample Size (ESS) of the prior distribution~\citep{morita_2008, neuenschwander_2020}, which can be interpreted as the amount of information contributed by the prior in terms of the equivalent number of patients in the Bayesian analysis.  

On the other hand, suppose some information about the control response is available (e.g., from historical trials or an earlier phase of the same study) and can be summarized by a distribution:  
\begin{equation}\label{hist.control}  
    \mu^H_C \sim \pi^H_C \left(m^H_C, l^H_C \right),  
\end{equation}  
where \( l^H_C \) represents either \( \sigma^H_C \) or \( \phi^H_C \), depending on the nature of the response variable. The quantity \( l^H_C \) is linked to \( \mathcal{I}^H_C \), which represents the amount of external information available, i.e., the ESS of the historical distribution. While this historical distribution could, in principle, be used as a prior, it may introduce issues: (i) heterogeneity between historical and concurrent control data could lead to bias in the control response estimation, and (ii) if the historical ESS is much larger than the control sample size in the trial, the prior may dominate the analysis, potentially compromising the integrity of the results.  

Suppose an interim analysis is planned at a given fraction of information \( t \in (0,1) \), after \( n^i_C + n^i_T = tN \) patients have been recruited, where \( n^i_C \) and \( n^i_T \) represent the number of patients in the control and treatment arms at the time of the interim analysis, respectively. Once their responses are observed, a measure of similarity \( \xi \) between historical and concurrent control data can be evaluated.  
Based on this similarity, a prior data-dependent distribution for the parameter \( \mu_C \) may be defined such that  
\begin{equation}\label{prior.control}
l^H_C = r(\xi, t, \mathcal{I}^i_C),
\end{equation}
where the function \( r(\cdot) \) determines the amount of information leveraged from the historical distribution. Here, \( \mathcal{I}^i_C \) represents the information available in the control arm at the interim analysis, and \( \xi \) quantifies the degree of heterogeneity between historical and current data at that time. This procedure preserves the mean value while modifying the ESS.

The trial is declared successful if  
\begin{equation}\label{eq:final}  
    P\left( \Delta > 0 \;\vert\; \{Y_C\}_{N_C}, \{Y_T\}_{N_T}, \pi^H_C, \pi^0_T \right) > \eta,  
\end{equation} 
where \( N_C > n^i_C \) and \( N_T > n^i_T \) (with \( N_C + N_T = N \)) represent the sample sizes in the control and treatment arms at the time of the final analysis. The threshold \( \eta \) is chosen to control the type I error rate at a predefined level \( \alpha \) under the null hypothesis.

\subsection{Distance criterion}\label{sec2.1}

At IA, we compute the interim posterior distribution for the control response $\mu_C$ based \textit{solely} on available concurrent data, with density  
\(
g_C^i \left(\mu_C \;|\; Y^i_C \right) \propto \mathcal{L}(Y^i_C \mid \mu_C) \pi^0_C(\mu_C),
\)
where $\pi^0_C(\mu_C)$ is the probability density function of a weakly informative prior distribution with mean $m^0_C$ and scale parameter $l^0_C$. A metric reflecting the degree of similarity between historical and concurrent control information is given by the Hellinger distance between $g^i_C$ 
and $\pi^H_C$, namely,
\begin{equation}\label{hellinger_generic}
    H \left( g^i_C, \pi^H_C \right) = \left[1 - \int  \sqrt{g^i_C(x) \pi^H_C(x)} \; dx\right]^{\frac{1}{2}}.
\end{equation}
For normal outcomes and normal prior distribution, Equation~\ref{hellinger_generic} can be written as
\begin{equation}\label{hellinger_normal}
    H \left( g^i_C, \pi^H_C \right) = \left[ 1-\sqrt{\frac{2 \sigma^i_C \sigma^H_C}{(\sigma^i_C)^2 + (\sigma^H_C)^2}} e^{-\frac{1}{4}\frac{\left(m^i_C - m^H_C \right)^2}{(\sigma^i_C)^2 + (\sigma^H_C)^2}}\right]^{\frac{1}{2}} ,
\end{equation}
while for binary variable and a Beta distribution as prior, it is given by:
\begin{equation}\label{hellinger_beta}
    H \left( g^i_C, \pi^H_C \right) =
    \left[ 1 - \frac{B\left(\frac{m^i_C \phi^i_C + m^H_C \phi^H_C}{2}, \frac{(1-m^i_C) \phi^i_C + (1-m^H_C) \phi^H_C}{2} \right)}
    {\sqrt{B(m^i_C \phi^i_C, (1-m^i_C) \phi^i_C) B(m^H_C \phi^H_C, (1-m^H_C) \phi^H_C)}} \right]^{\frac{1}{2}},
\end{equation}
where $B(a, b)$ is the Beta function:
\(
    B(a, b) = \int_0^1 z^{a-1} (1-z)^{b-1} dz.
\)
From Equations~\ref{hellinger_normal} and \ref{hellinger_beta}, it follows that $H=0$ if and only if $m^i_C = m^H_C$ and $l^i_C = l^H_C$, where \( l^H_C \) represents either \( \sigma^H_C \) or \( \phi^H_C \), depending on the nature of the response variable.
In addition, in case of continuous outcome, $H$ is an increasing function of the variables $d = m^i_C - m^H_C$
and $s =\big|l^i_C - l^H_C\big|$, so that $H \to 1$ only if $|d| \to \infty$ or $s \to \infty$. While this result can be straightforwardly shown using Equation~\ref{hellinger_normal}, it is not true in general for the binary outcome. Hereafter, we refer to $d$ as \textit{drift}.
In our context, we aim to preserve the quantity of information given by both $l^i_C$ and $l^H_C$, and given that $l^i_C \neq l^H_C$ in general, it follows that the minimum value $H^{\mbox{min}}$ of $H$, when $l^i_C$ and $l^H_C$ are considered as constants, is reached when $d=0$ in the continuous case, while it needs to be approximated computationally in the binary case. If the prior distribution arises as a mixture, as is common with a meta‐analytic prior, let
\(
\pi^H_C\sim q(x)\),
\(q(x) = \sum_{k=1}^K w_k\,q_k(x),
\)
where each \(q_k(x)\) is, for example, a normal density in the continuous case or a beta density in the binary case, and the \(w_k\) are mixture weights. Then \(H^{\mathrm{min}}\) can be computed via numerical optimisation.
Then, a new criterion derived from the Hellinger distance can be defined as
\begin{equation}\label{Hellinger_normalized}
H^*\left(g^i_C, \pi^H_C \right) = \frac{H\left(g^i_C, \pi^H_C \right) - H^{\mbox{min}}}{1-H^{\mbox{min}}}.
\end{equation}

\begin{figure}[tb]
\begin{center}
  \includegraphics[width=\textwidth]{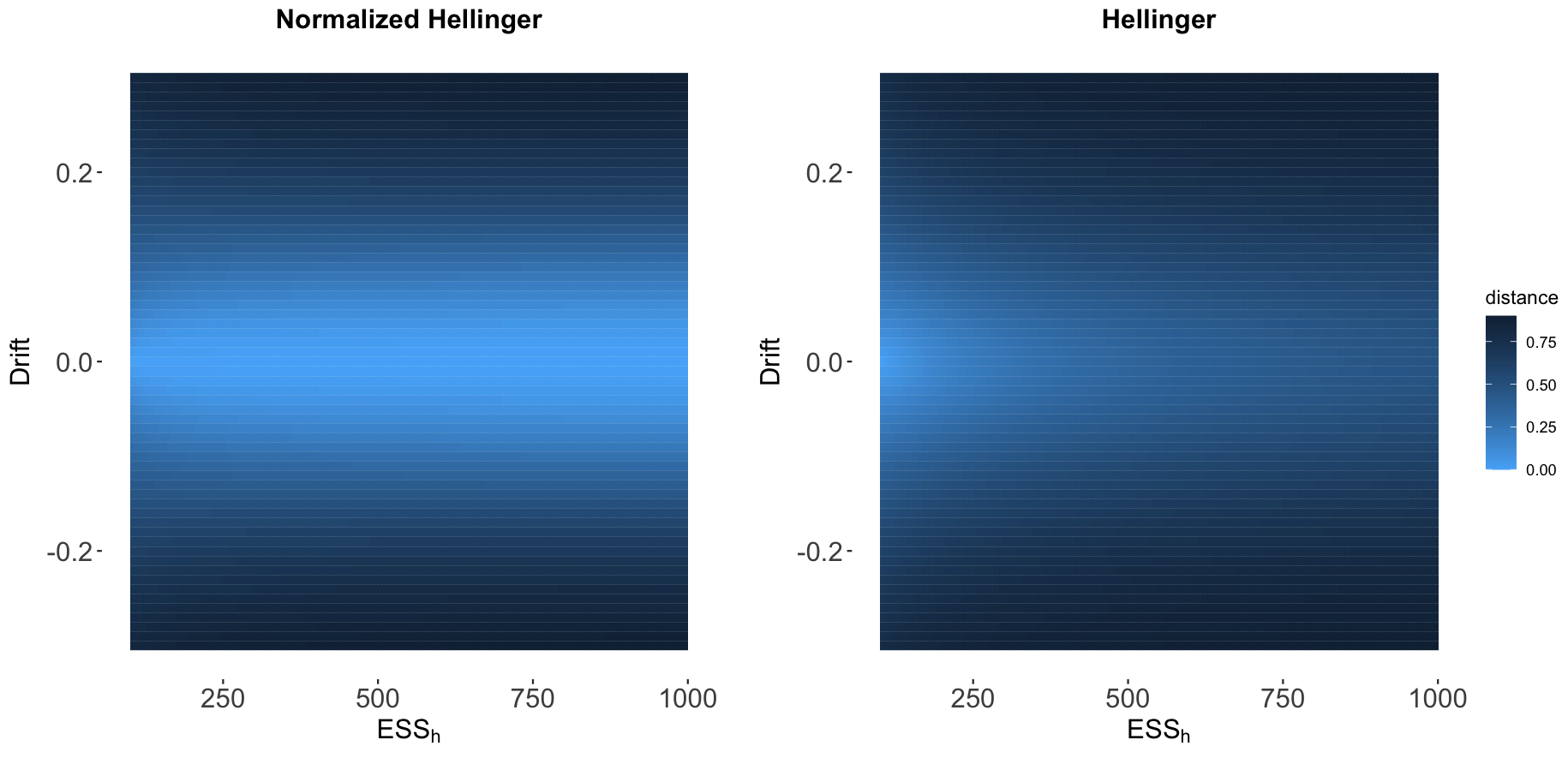}
  \caption{Comparison between the Normalized Hellinger Distance $H^*$ (left panel) versus the Hellinger distance $H$ (right panel) depending on the drift $d$ and the effective sample size (ESS) of the historical control distribution $\pi^H_C(x)$. A posterior distribution $g_C(x)$ is used assuming 50 patients are recruited in the control arm. }\label{fig:distances}
\end{center}
\end{figure}

This new criterion is less sensitive to differences in the information content of the two distributions. As shown in Figure~\ref{fig:distances}, it remains stable when varying the informativeness of the prior distribution, quantified in terms of ESS. In this manuscript, we employ the ELIR ESS of \cite{neuenschwander_2020}. Alternative methods to account for differing ESSs in the Hellinger distance build a new distribution by weighting the likelihood of the more informative distribution and then computing the Hellinger distance between this weighted distribution and the other distribution~\citep{ollier_2020,calderazzo_2024}. Although both the weighting and minimum-distance approaches share the empirical Bayes philosophy, the former is generally more computationally intensive.

\subsection{Determination of the similarity parameter}\label{sec2.2}
We set
\begin{equation}\label{eq:xi}
  \xi = f\bigl(1 - H^*\bigr)\,\mathbbm{1}\{H^* \le \gamma\},
\end{equation}
where $f(x)$ is an increasing function of $x$, $\mathbbm{1}\{\cdot\}$ is the indicator function (equal to 1 if its argument is true, and 0 otherwise), and $\gamma$ is a tuning parameter chosen before the start of the trial. 
The function $f$ controls the degree of similarity preserved from the raw computed similarity, $1 - H^*$. A natural choice is the identity function,
\(
f(x) = x,
\)
so that $f(1 - H^*) = 1 - H^*$. 
The indicator function serves as an Occam’s razor, preventing borrowing when the historical and current data are deemed insufficiently similar. The threshold parameter $\gamma$ specifies the maximum allowable discrepancy. The impact of $\gamma$ on $\xi$ will be evaluated in the simulation study.

\subsection{Two stage $\xi$-driven adaptive designs}\label{sec:two-stage}

Borrowing information from external controls may be beneficial in reducing the sample size of the concurrent trial; on the other hand it may lead to undesirable increase in type I error in case of substantial heterogeneity between external and concurrent information.
In order to mitigate such situation, we propose a dynamic sample size reassessment, where the number of patients to be recruited is dynamically varied throughout the study based on the degree of ``exchangeability" between historical and concurrent information. 
Suppose that at IA, the posterior distribution $g^i_C$ for $\mu_C$ is computed and the parameter $\xi$ is determined as described in Section \ref{sec2.2}. In order to optimize the second stage randomization in the trial, two different strategies are proposed.

\subsubsection{Sample Size Reassessment}\label{2.3.1}
In the first strategy (called design 1 from now on), the second stage sample size in the control arm may be adjusted by setting
\begin{equation}\label{SS_control}
    n^{(2)}_C = \left\lfloor\left(1-t\right)\left(1- \frac{\xi}{\lambda} \right)\frac{N}{R+1}\right\rfloor,
\end{equation}
where the symbols $\left\lfloor x \right\rfloor$ denote the floor function of the real number $x$. The second stage sample size in the treatment arm is kept $n^{(2)}_T = \left(1-t\right)\frac{RN}{R+1}$. 
This choice guarantees that when there is full heterogeneity between concurrent and historical controls (i.e. $\xi \rightarrow 0$) the second stage sample size for control arm is $ n^{(2)}_C=\left(1-t\right)\frac{N}{R+1} $, the allocation ratio $R:1$ is maintained. 
On the contrary, when full homogeneity between concurrent and historical controls is observed at the time of the interim analysis (i.e. $\xi \rightarrow 1$), then $ n^{(2)}_C = \left\lfloor\left(1-t\right)\left(1- \frac{1}{\lambda} \right)\frac{N}{R+1}\right\rfloor $ and accordingly the second stage allocation ratio becomes $R^*:1$ , where $R^* \approx \frac{\lambda R}{\lambda - 1}$. Notice that setting $\lambda=1$ leads to a loss of randomization in case of $\xi \rightarrow 1$ (all second stage patients are allocated to the treatment arm); hence setting $\lambda > 1$ can be considered as it draws an upper bound for the allocation ratio. 
Using this strategy, the total sample size is reduced by $n_s \approx (1-t)\frac{\xi N}{\lambda(R+1)}$ patients. 

\subsubsection{Allocation Ratio Imbalance}\label{2.3.2}
Using this strategy (called design 2 from now on) a total sample size of N is maintained, however a randomization imbalance towards the treatment arm is induced by placing $(1-t)\frac{\xi N}{\lambda(R+1)}$ patients, who were planned to be initially allocated to the control arm, to the treatment arm.
Indeed, the second stage sample size in the control arm may be adjusted as in Equation \ref{SS_control} while the second stage sample size in the treatment arm is varied to 
\begin{equation}\label{SS_treatment}
    n^{(2)}_T \approx \left(1-t\right)\left(R + \frac{\xi}{\lambda} \right)\frac{N}{R+1}
\end{equation}
As in the previous strategy, when $\xi =0$ the pre-planned second stage sample sizes are maintained as well as the allocation ratio to $R:1$. However, when full homogeneity between concurrent and historical controls is observed at the time of the interim analysis ($\xi \rightarrow 1)$, then $ n^{(2)}_C \approx \left(1-t\right)\left(1- \frac{1}{\lambda} \right)\frac{N}{R+1} $, $ n^{(2)}_T \approx \left(1-t\right)\left(R + \frac{1}{\lambda} \right)\frac{N}{R+1} $ and the allocation ratio becomes $R^*:1$, where $R^* \approx \frac{\lambda R + 1}{\lambda - 1}$. 

\subsubsection{Prior distributions for control arm}\label{2.3.3}

In both settings, at the end of the trial, we adjust $\pi^H_C$ so that its ESS equals the number of patients no longer randomized to the control arm,
\(
n_s 
\)
In the Gaussian case, if \(n_s>0\) we scale the prior standard deviation \(\sigma_{C}^{H}\) by the factor \(\sqrt{n_s/n_p}\), i.e.\ setting
\(
\sigma_{C, new}^H \;=\; \frac{\sigma_C^H}{\sqrt{n_s/n_p}},
\)
where $n_p$ is the initial ESS of $\pi^H_C$ from,
and in a mixture‐of‐normals prior we instead find, via a simple optimisation function, \(v>0\) such that dividing each component’s standard deviation by \(v\) yields \(\mathrm{ESS}(\pi^H_{C, new})=n_s\). In the binary case, if \(n_s>0\) we multiply the precision parameter \(\phi^H_C\) by 
\(n_s/n_p\)
i.e.\ setting
\(
\phi^H_{C, new} \;=\; \phi^H_C \times 
\frac{n_s}{n_p}\),
and for a mixture of Betas we choose \(v>0\) so that multiplying each \(\phi\) by \(v\) yields \(\mathrm{ESS}(\pi^H_{C, new})=n_s\). If \(n_s=0\), the same adjustments apply with \(\mathrm{ESS}(\pi^H_{C, new})\) forced to 1. For simplicity, and without appreciable loss of performance, one may simply adopt a non-informative \(\mathrm{Beta}(0.5,0.5)\) prior in the binary scenario.

\subsection{On the specification of the design parameters}
In the designs presented in Sections \ref{2.3.1} and \ref{2.3.2}, several parameters require pre-trial specification: $t$, corresponding to the IA information fraction in term of sample size; and $\gamma$ and $\lambda$, which govern information borrowing. 
A rationale for selecting $\lambda$, ensuring an upper bound for the second-stage allocation ratio $R^*$, was provided in Sections \ref{2.3.1} and \ref{2.3.2}.  Alternative approaches could involve specifying $\lambda$ to guarantee a minimum number of second-stage patients in the control arm. 

Regarding $t$ and $\gamma$, the selection can be proposed based on a probability statement on the distribution of $\xi$. Specifically, a joint specification of these parameters can be achieved by constraining the probability of borrowing under a specific level of drift $\delta^*$ :
\begin{equation}
\label{prob_dist}
    \mathbb{P}\left( \xi > 0 \; \big{|} \; \delta^* \right) < \varepsilon \; ,
\end{equation}
where $\delta^*$ represents a maximum acceptable drift (hereinafter referred to as MAD) between concurrent and historical data, and $\varepsilon$ is a small probability (e.g.before 20\%, say). The MAD should ideally represent that level of departure from historical assumption under which only a small probability of borrowing is allowed, and must be carefully discussed with the clinical team.

Parameters $t$ and $\gamma$ influence Equation~\ref{prob_dist} in different manners.  Larger values of $t$ reduce the variance of the $\xi$ distribution, improving estimation precision and facilitating satisfaction of the criterion in Equation~\ref{prob_dist}.  Conversely, $\gamma$ directly affects the definition of $\xi$, with larger values of $\gamma$ leading to stochastically smaller $\xi$ values, again facilitating satisfaction of Equation~\ref{prob_dist}. 

In principle, increasing $t$ or decreasing $\gamma$ yields a similar influence on
Equation~\ref{prob_dist}, resulting in multiple pairs $(t,\gamma)$ that satisfy the
prescribed constraint. To obtain a unique parameter configuration, we introduce an
optimality criterion based on the expected number of second--stage patients who are
either not allocated to control (if design 1 is employed) or reallocated from the control to the treatment arm (if design 2 is employed). Specifically, we propose selecting the pair $(t,\gamma)$ that maximizes this quantity. An example of the procedure in a practical setting is given in Section \ref{sec4}.
We note that although $\gamma$ may be freely specified, the timing of the interim
analysis, and hence the value of $t$, is frequently constrained by practical and
operational considerations beyond statistical principles. Consequently, alternative
strategies may be adopted; for example, one may fix $t$ at the minimum admissible
information fraction for the interim analysis and subsequently adjust $\gamma$ to
satisfy Equation~\ref{prob_dist}.

\section{Simulation Study}\label{sec3}
A simulation study was performed to assess the performance of the proposed designs under different configurations. Specifically, the two different designs proposed in Sections \ref{2.3.1} and \ref{2.3.2} are considered; namely, design 1 with sample size re-estimation based on $\xi$ at the IA, and design 2 with fixed sample size and a reassessment of the allocation ratio based on $\xi$ at the IA. We present the results for the continuous case with a ``single" (nor mixture) prior; full results for mixture priors and binary outcomes are provided in Appendix.

\subsection{Setting}\label{sec3.1}
In both designs, patients' normal responses $\mbox{Y}_T$ and $\mbox{Y}_C$ are considered in both arms, with unit variance and means respectively equal to the true parameters $\theta_T$ and $\theta_C$. An initial total sample size of \( N = 200 \) is considered, motivated by a frequentist sample size calculation based on a Cohen's \( d = 0.4 \) (and the previously mentioned unit variance), unilateral \(\alpha\) of 0.025 and power of 0.80. 
An informative distribution for the control parameter $\pi^{H}_C$ is set, centered in zero and worth 70 patients ($n_p= 70$) and a non-informative and conservative prior distribution for the treatment parameter, centered in zero and worth one patient. 
Different  ``timings" for the IA are explored, namely $t \in \{0.3, 0.4, 0.5, 0.6 \}$ and equal randomization of the patients across the two arms is considered for the first stage, so that $R^{(1)}=1$. 
To compute the minimal distance, we use a Gaussian distribution with mean equal to that of the prior distribution and standard deviation equal to the unit standard deviation divided by the square root of the sample size at IA. $g_C^i \left(\mu_C \;|\; Y^i_C \right)$ uses the same standard deviation, but its mean is estimated from the actual control data. In the case of a mixture prior, the standard deviation is defined as before, while the mean is determined via an optimization algorithm. An approximation, using the prior mean, is also tested; in this case, all negative distances are set to 0 (results are available in Appendix).
In the main scenarios, the parameter $\xi$ is used for the assessment of the similarity between concurrent and historical control information, with $\gamma=0.3$, set in order to enable borrowing only in case $H^{*}(g_C^{i},\pi^{H}_C) < 0.3$. We also set $f$ to be the identity function in Eq.~\ref{eq:xi}. Finally the parameter $\lambda$ is set to 1. Sensitivity analyses on this choices are then performed.

As a comparator, we use the analysis that would have been performed in the absence of interim analyses, with non-informative prior distributions, that is, priors with means set to 0 and effective sample sizes of 1 for both arms. A total of 5000 trials are simulated for each set of parameters. While \( \pi_C^H \) remains constant, \( \theta_T \) and \( \theta_C \) vary according to the explored drift $d$ and hypothesis (i.e., whether computing the type I error or the power).
The metrics assessed include the change in type I error (when $\theta_T = \theta_C$) and power (when $\theta_T - \theta_C = 0.4$) relative to the comparator, bias (using the posterior mean as the point estimate for \( \Delta \)), the number of patients saved (or re-allocated, depending on the design employed), and the length of the 95\% (equal tails) credible interval. The posterior mean is used as point estimator and we set $\eta = 0.975$ in Eq.\ref{eq:final}, inspired by the frequentist approach. All analyses are performed using R software version 4.3.0, with the RBesT package version 1.7.3 for the Bayesian analysis and ESS computation.

\subsection{Results}\label{sec3.2}

\begin{figure}[tb]
\begin{center}
  \includegraphics[width=0.9\textwidth]{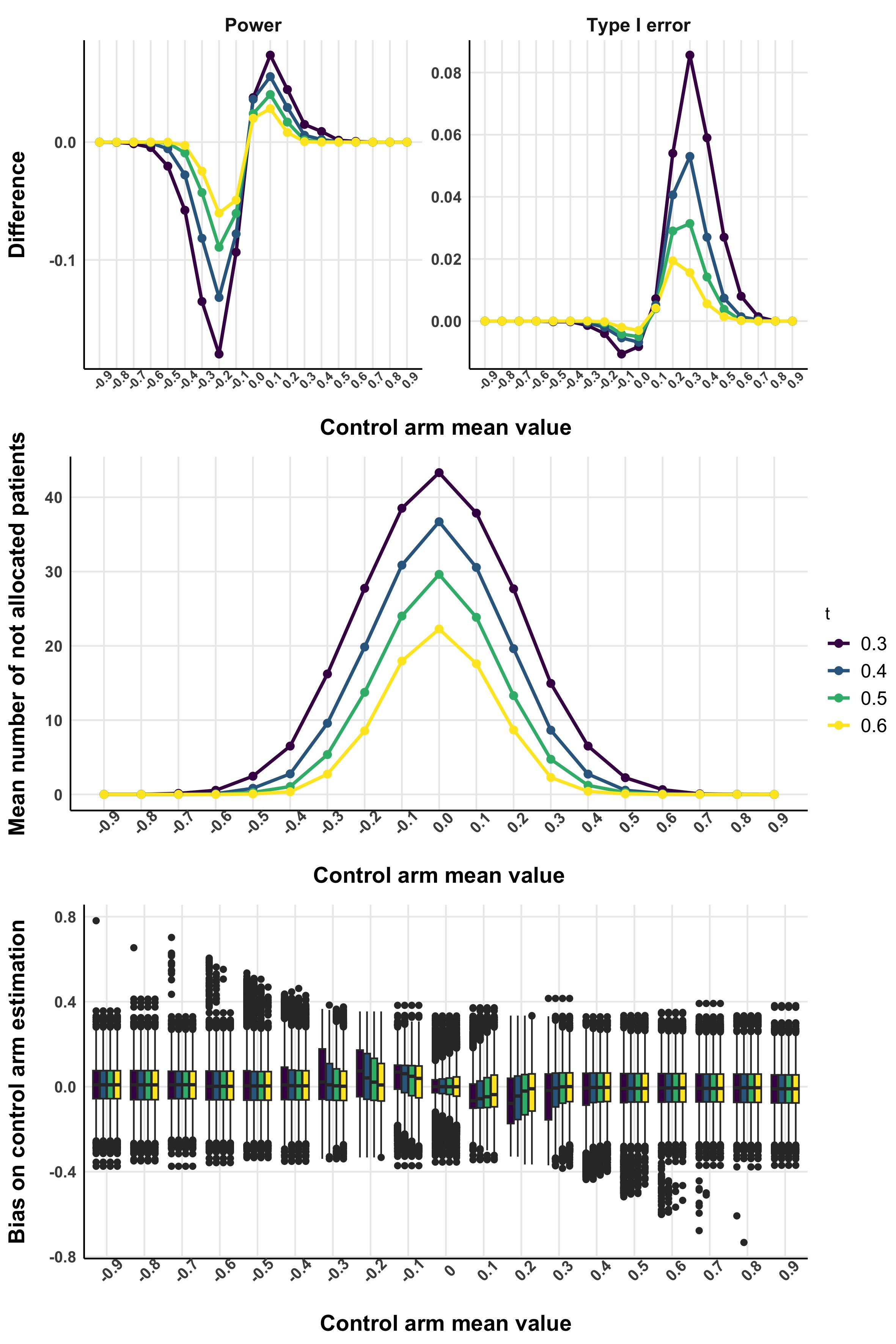}
  \caption{Results for the first type of design in terms of power (top left panel), type I error (top right panel), mean number of not allocated patients (middle panel) and control arm estimation bias (bottom panel) at different information fractions $t$.}\label{fig:resmain}
\end{center}
\end{figure}

Figure~\ref{fig:resmain} illustrates the effect of varying trial control arm mean values $\theta_C$ on statistical performance metrics, under different  fraction of information \( t \in \{0.3, 0.4, 0.5, 0.6\} \) for design 1.
The top left panel shows the difference in power relative to the comparator. A loss in power is observed for negative drifts in the control arm mean (i.e., when the control arm under-performs expectations and the prior distribution is then too conservative), particularly for smaller drift, with losses of approximately -0.18, -0.13, -0.09 and -0.06 for $t$ equals to  0.3, 0.4, 0.5 and 0.6, respectively, reached at $d = -0.2$. As the control mean increases toward zero and beyond, the power recovers and even slightly exceeds that of the comparator for positive drift scenarios. Indeed, the gain in power reaches 0.07 when $d = 0.1$ and $t=0.3$.

The top right panel of Figure~\ref{fig:resmain} reports the difference in type I error. A clear inflation of type I error occurs when the control arm mean is positively biased, peaking around $d = 0.3$. This inflation is more severe for smaller effect sizes, especially when \( t = 0.3 \) with an increase of 0.09, and diminishes with increasing \( t \). For negative drifts, a modest reduction in type I error is observed (until -0.01 for $d = -0.1$ and $t = 0.3$), and therefore it is still controlled.

The middle panel of Figure~\ref{fig:resmain} displays the mean number of not allocated patients as a function of the control arm mean. The number of unallocated patients is maximum when the control arm mean coincides with the one of the prior, 
going from 22 for $t = 0.6$ to 43 for $t=0.3$, while symmetrically decreasing as the control mean differs from the one of the prior. As expected, values decrease as $t$ increase.
Figure~\ref{fig:resmain} bottom displays the bias in control arm estimation, defined as the difference between the estimated and true control means, and shows that while the bias remains relatively centered around zero across most scenarios, a systematic underestimation emerges when the control mean is positively biased (especially beyond 0.3), with wider variability and some degree of overestimation seen for large negative control means.

\begin{figure}[tb]
\begin{center}
  \includegraphics[width=0.9\textwidth]{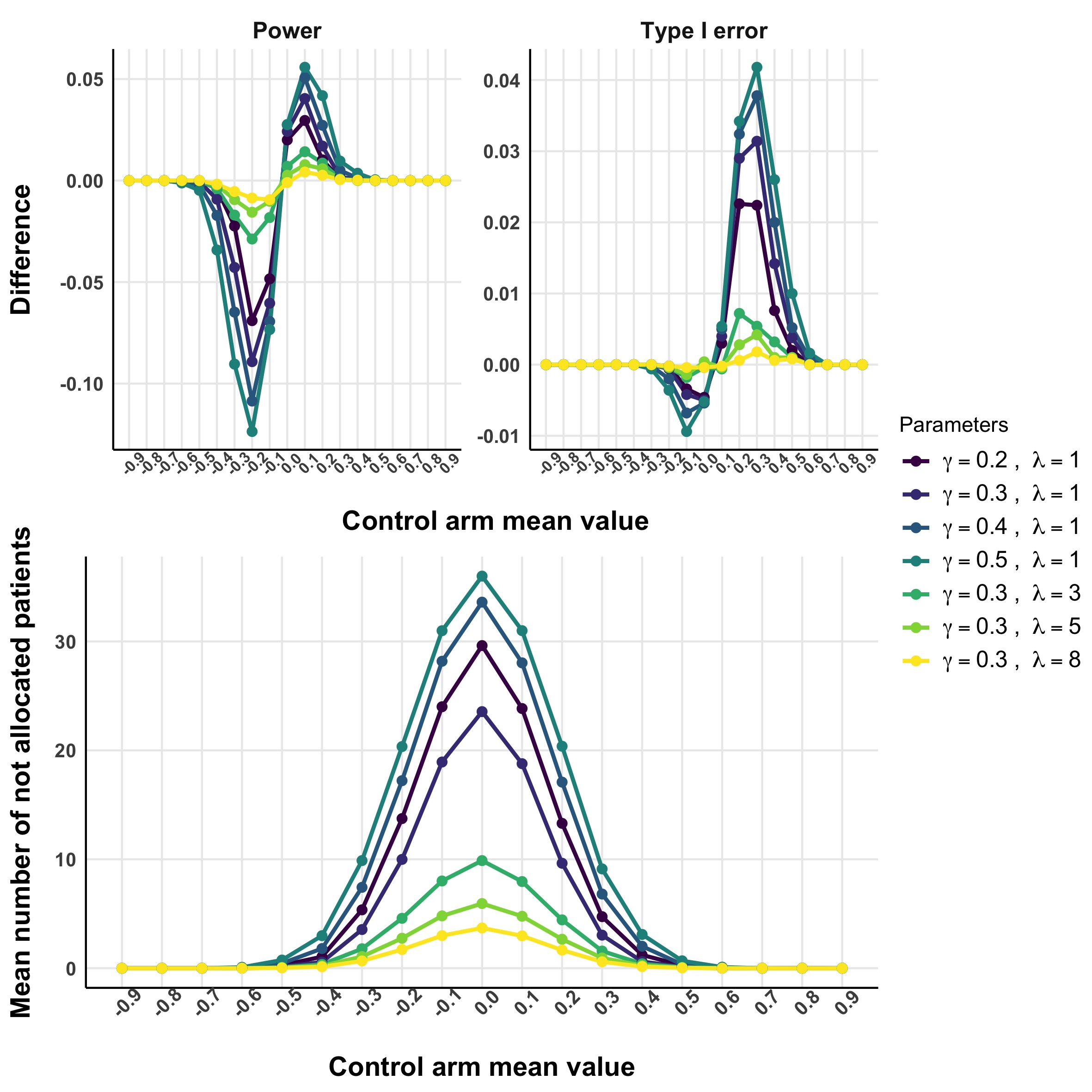}
  \caption{Results for the first type of design in terms of power (top left panel), type I error (top right panel) and mean number of not allocated patients (bottom panel) for $t=0.5$ when varying $\gamma$ and $\lambda$.}\label{fig:ressens}
\end{center}
\end{figure}

Figure~\ref{fig:ressens} presents the sensitivity‐analysis results for varying values of \(\gamma\) and \(\lambda\) at \(t = 0.5\). The base design \((\gamma=0.3,\ \lambda=1)\) is plotted alongside each alternative.
As \(\gamma\) increases, both power and 
type I error rate rise under positive drift (until 0.06 and 0.04 at $d$ equals to 0.1 and 0.3, respectively, for $\gamma = 0.5$) and fall under negative drift (until respectively \{-0.12, -0.009\} at $d \in \{-0.2, -0.1\}$ for $\gamma = 0.5$). Conversely, increasing \(\lambda\) (i.e.\ allocating a larger proportion of patients to the control arm) attenuates the impact of drift on both power and type I error but substantially reduces the number of patients \(n_s\) assigned to the experimental arm (going from 30 to 4 patients for respectively $\lambda \in \{1, 8\}$ at $d = 0$).

\begin{figure}[tb] 
\begin{center}
  \includegraphics[width=0.9\textwidth]{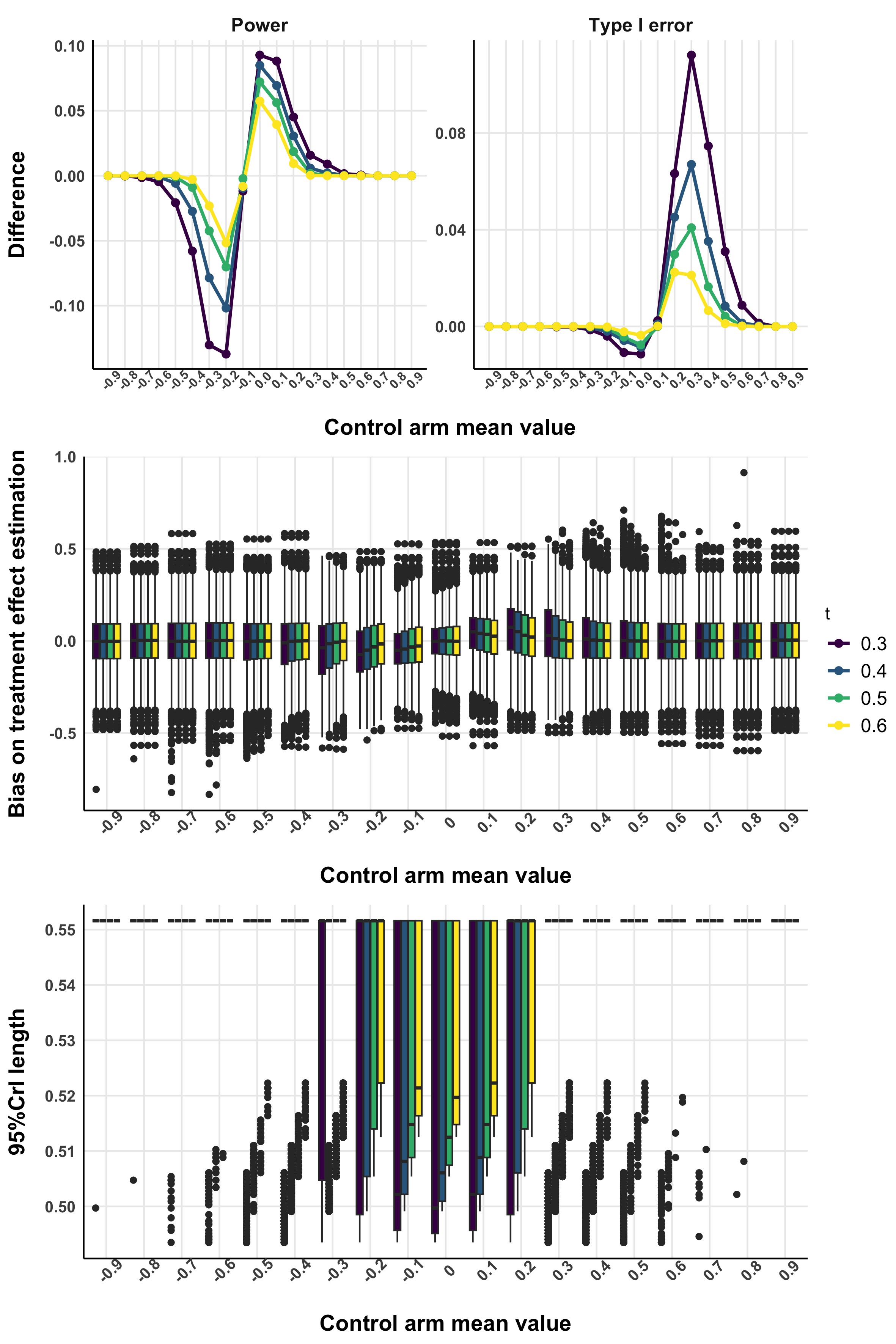}
  \caption{Results for the second type of design in terms of power (top left panel), type I error (top right panel), treatment effect estimation bias (middle panel) and 95\% credible interval length (bottom panel) at different information fractions $t$.}\label{fig:resdes2}
\end{center}
\end{figure}

Figure~\ref{fig:resdes2} presents the results for the second design.  The top panel reproduces the sensitivity patterns seen in design 1. The middle panel displays the bias in the treatment–effect estimate, rather than just control–arm bias. The bottom panel depicts the length of the 95\% credible interval: it is widest for substantial negative drifts, contracts as the true control mean approaches the prior mean, and then levels off for large positive drifts.

Full details regarding the binary case, the use of mixture priors for the normal cases and the proposed approximation to compute the minimum Hellinger distance are provided in the Appendix. The results are consistent with those shown in this manuscript, and the approximation produces no differences in the results.

\section{Practical application}\label{sec4}

To illustrate how this method can be applied at the planning stage of a clinical trial, we consider the study described in \cite{hueber_2012} (registered at ClinicalTrials.gov, NCT01009281), which used a Bayesian analysis based on the MAP prior.
In particular, since the previous simulations focus on classical operational characteristics, in this section we show how using the maximum acceptable drift defined in Eq.~\ref{prob_dist} and the potential number of patients saved at the trial planning stage.


In the original clinical trial, which assessed the efficacy of AIN457 in Crohn’s disease using the primary endpoint, change in the Crohn’s Disease Activity Index (CDAI) at six weeks, the sample size was set at 60 patients, 40 for the treatment arm and 20 for the control arm and the equivalent of 20 patients from historical control data were included in the analysis to balance the sample size in the two arms. 

For the purpose of understanding how the proposed methodology could be implemented in practice, let us retrospectively suppose that no borrowing is allowed initially, so that the sample size per arm is $N_T = N_C = 40$ patients ($R = 1$); while the second stage sample size is adapted at the time of an interim analysis (based on the similarity between concurrent and historical control data $\xi$) in order either to discount the number of patients allocated to the control arm (design 1), or re-allocate these patients to the treatment arm (design 2).  

We further assume that the control response, $Y_C$, follows a normal distribution with mean $\mu_C$ and standard deviation $\sigma_C = 88$, reflecting the prior hypothesis rather than the estimate obtained at the end of the trial.
As in the original study, data information from six previous historical trials (including 671 individuals) is transformed into a prior distribution by means of the MAP methodology using $m_{C}^{0,H} \sim N(0, (\sigma_C^{H})^2)$ (conservative)  and $\sigma_{C}^{0,H} \sim N^+(0, (\sigma_C^{H}/2)^2)$ (very conservative). To compute the ESS of the MAP prior, we assume a standard deviation of 88 for the change from baseline, consistent with the hypothesis on $Y_C$. For our case study, we obtain as mean of the prior distribution approximately -50, with a standard deviation of 18. A visual representation of the MAP can be found in the Appendix. 

In Table \ref{tab:cont_case_study_res}, we illustrate how Eq.~\ref{prob_dist} can guide the selection of hyperparameters during the design phase of a hypothetical trial. Specifically, for several information fractions $t \in \{0.4, 0.5, 0.6\}$ and choices of the parameter $\gamma \in \{0.2, 0.3, 0.4, 0.5\}$, we evaluate the posterior probability that the similarity parameter $\xi$ exceeds zero under a maximum acceptable drift $\delta^*$. Because $\delta^*$ should ultimately be defined in consultation with the clinical team, simulations were conducted under a range of plausible values $\delta^* \in \{\pm 20, \pm 30, \pm 40\}$. Due to the small sample size of the trial, no cap to the second stage allocation ratio is forced, so that the parameter $\lambda$ is set to 1.

The results indicate that, for fixed $\gamma$ and $\delta^* \neq 0$, increasing the information fraction $t$ reduces the borrowing probability under MAD, meaning that less information is expected to be borrowed from the MAP prior, thus making the inference be primarily guided by concurrent data. This pattern arises because larger $t$ yields a more precise values for the similarity parameter $\xi$, which appropriately reflects low accordance under drift conditions.

For fixed $t$ and $\delta^*$, the results also show that larger values of $\gamma$ correspond to increased borrowing. This follows from Eq.~\ref{eq:xi}, where $\gamma$ defines the threshold for the normalized Hellinger distance below which borrowing occurs (i.e., when $\xi > 0$).

Holding $t$ and $\gamma$ constant, greater values of the MAD  $|\delta^*|$ result in lower borrowing probabilities, whereas smaller MAD increase the likelihood of borrowing. Although the choice of $\delta^*$ should primarily reflect clinical views on plausible endpoint variability, it should also be coherent with the information encoded in the MAP prior. In our setting, for example, deeming a drift of $|\delta^*|=20$ as unacceptable might be questionable, as it would imply limited confidence in the MAP prior itself (since the PDF of the MAP in the points $\delta^*=20$ and $\delta^*=-20$ exhibits large values). 

Let us suppose that the clinical team identifies $\delta^* = 40$ points as the MAD for the CDAI on the control arm, and assume that a borrowing probability of $\varepsilon = 15\%$ is considered acceptable under this scenario, Table~\ref{tab:cont_case_study_res} indicates that the parameter combinations for which Eq.~\ref{prob_dist} is satisfied are: $\left(t,\gamma\right) \in \{(0.4, 0.2), (0.5, 0.2), (0.6, 0.2), (0.6, 0.3)  \}$.
All these couples allow for a borrowing probability of less than 15\% under the critical drift $\delta^*$. In order to select uniquely the pair $\left(t,\gamma\right)$ we propose to choose the configurations which maximize the number of second stage patients not-allocated (or re-allocated to the treatment arm). Among the four pairs found, the pair $t=0.4$ and $\gamma=0.2$ is the one which allows for a maximum benefit, with 13 patients saved (versus 11.9 and 10, respectively for the other three configurations). Following this, the latter combination of parameter is suggested for the current design.  

\begin{table}
\caption{Probability that $\xi$ is greater to 0 given $\delta^*$ and mean number of saved patients for several values of $\gamma$ and several values of $t$ for the practical application.}
\label{tab:cont_case_study_res}
\begin{center}
\scriptsize
\begin{tabular}{l|lll|lll|lll|lll}
\hline
& \multicolumn{3}{c|}{$\gamma = 0.2$} & \multicolumn{3}{c|}{$\gamma = 0.3$} & \multicolumn{3}{c|}{$\gamma = 0.4$} & \multicolumn{3}{c}{$\gamma = 0.5$}\\ 
\hline
& $t = 0.4$ & $t = 0.5$ & $t = 0.6$ & $t = 0.4$ & $t = 0.5$ & $t = 0.6$ & $t = 0.4$ & $t = 0.5$ & $t = 0.6$ & $t = 0.4$ & $t = 0.5$ & $t = 0.6$ \\ \hline
& \multicolumn{12}{c}{$\mathbb{P}\left( \xi > 0 \; \big{|} \; \delta^* \right)$} \\
 \hline
$\delta^* \approx -40$ & 0.15 & 0.10 & 0.08 & 0.23 & 0.17 & 0.13 & 0.33 & 0.26 & 0.21 & 0.45 & 0.38 & 0.32 \\
$\delta^* \approx -30$ & 0.27 & 0.22 & 0.19 & 0.38 & 0.33 & 0.28 & 0.50 & 0.45 & 0.39 & 0.63 & 0.58 & 0.53 \\
$\delta^* \approx -20$ & 0.41 & 0.38 & 0.35 & 0.54 & 0.51 & 0.48 & 0.67 & 0.64 & 0.60 & 0.78 & 0.75 & 0.73\\
$\delta^* \approx 0$ & 0.58 & 0.57 & 0.58 & 0.72 & 0.72 & 0.72 & 0.83 & 0.83 & 0.83 & 0.91 & 0.91 & 0.91 \\
$\delta^* \approx 20$ & 0.40 & 0.37 & 0.35 & 0.53 & 0.49 & 0.47 & 0.65 & 0.61 & 0.59 & 0.76 & 0.73 & 0.71 \\
$\delta^* \approx 30$ & 0.26 & 0.21 & 0.18 & 0.37 & 0.31 & 0.27 & 0.48 & 0.42 & 0.38 & 0.60 & 0.55 & 0.51 \\
$\delta^* \approx 40$ & 0.14 & 0.10 & 0.07 & 0.22 & 0.16 & 0.12 & 0.31 & 0.24 & 0.19 & 0.43 & 0.35 & 0.29 \\ \hline
& \multicolumn{12}{c}{Mean number of not allocated patients} \\ \hline
$\delta^* \approx 0$ & 13 & 11 & 9 & 15 & 13 & 10 & 17 & 14 & 12 & 18 & 15 & 12 \\
\hline
\end{tabular}
\end{center}
\end{table}

\section{Conclusion}\label{sec5}

We proposed a Bayesian framework, based on the ``test and pool" approach, which can either reduce the sample size (design 1) or increase the proportion of patients allocated to the treatment arm (design 2) when the prior distribution for the control arm aligns with the interim trial data. As expected, while this approach can reduce the sample size and increase the power, it may also increase the type I error rate in determinate settings, consistent with findings from other Bayesian borrowing studies~\citep{kopp_2020,kopp_2024}. However, we demonstrated that the design parameters can be calibrated to achieve the desired operating characteristics, which can be extended to novel metrics~\citep{best_2025}, and mitigate the risks of power loss and type I error inflation. 

In this work, we measured the degree of heterogeneity between the historical and current control information at the interim analysis using a metric derived from the Hellinger distance. This choice was made for simplicity, as it allows for closed-form solutions in certain cases and provides a bounded measure, which enhances the interpretability of the results. However, other measures, such as the Kullback–Leibler or Jensen–Shannon divergence, could also be employed, possibly along with numerical approximations.

Our methodology only requires a prior distribution on the control response to summarize historical evidence. It is independent of the process by which this prior was obtained. Therefore, the method can be applied once the two populations, the external and the trial populations, have been shown to be homogeneous with respect to baseline covariates, or when the prior distribution is derived from other statistical methods, for example, from a propensity score–matched population~\citep{guo_2024, wang_cao_yao_2025}.

In addition, by examining the expected behavior of the distance measure in key scenarios, clinicians can also assess the feasibility of conducting a trial with a small sample size, where Bayesian inference is most commonly used.

The central feature of the proposed methodology is the incorporation of an 
interim analysis to adapt the second--stage randomization process. Consequently, the approach is particularly suitable when the time elapsed between randomization and outcome assessment is short, that is, when the interim analysis 
can be conducted shortly after recruitment of the final patient in stage~1. It is important to note that in settings where this condition does not hold, such as in the presence of delayed outcomes (e.g., Overall Survival, Progression--Free Survival), the methodology may not be appropriate, as the interim analysis may occur only after a substantial number of patients in stage~2 have already been recruited. Extensions of the proposed framework to accommodate such situations are left for future work.

\paragraph{Acknowledgements} 
The work of Moreno Ursino was supported by a grant from Inserm and the French Ministry of Health (MESSIDORE 2022, reference number Inserm-MESSIDORE N° 94). Sandrine Boulet and Moreno Ursino made equal contributions and are co-last authors. Natural language processing tools were used during manuscript preparation to assist with grammar and sentence structure.

\bibliography{historical_information_in_phase_II_trials_20250522_arXiv_version}%

\clearpage 

\appendix

\section{Mixture} 

\begin{figure}[h] 
\begin{center}
  \includegraphics[width=0.9\textwidth]{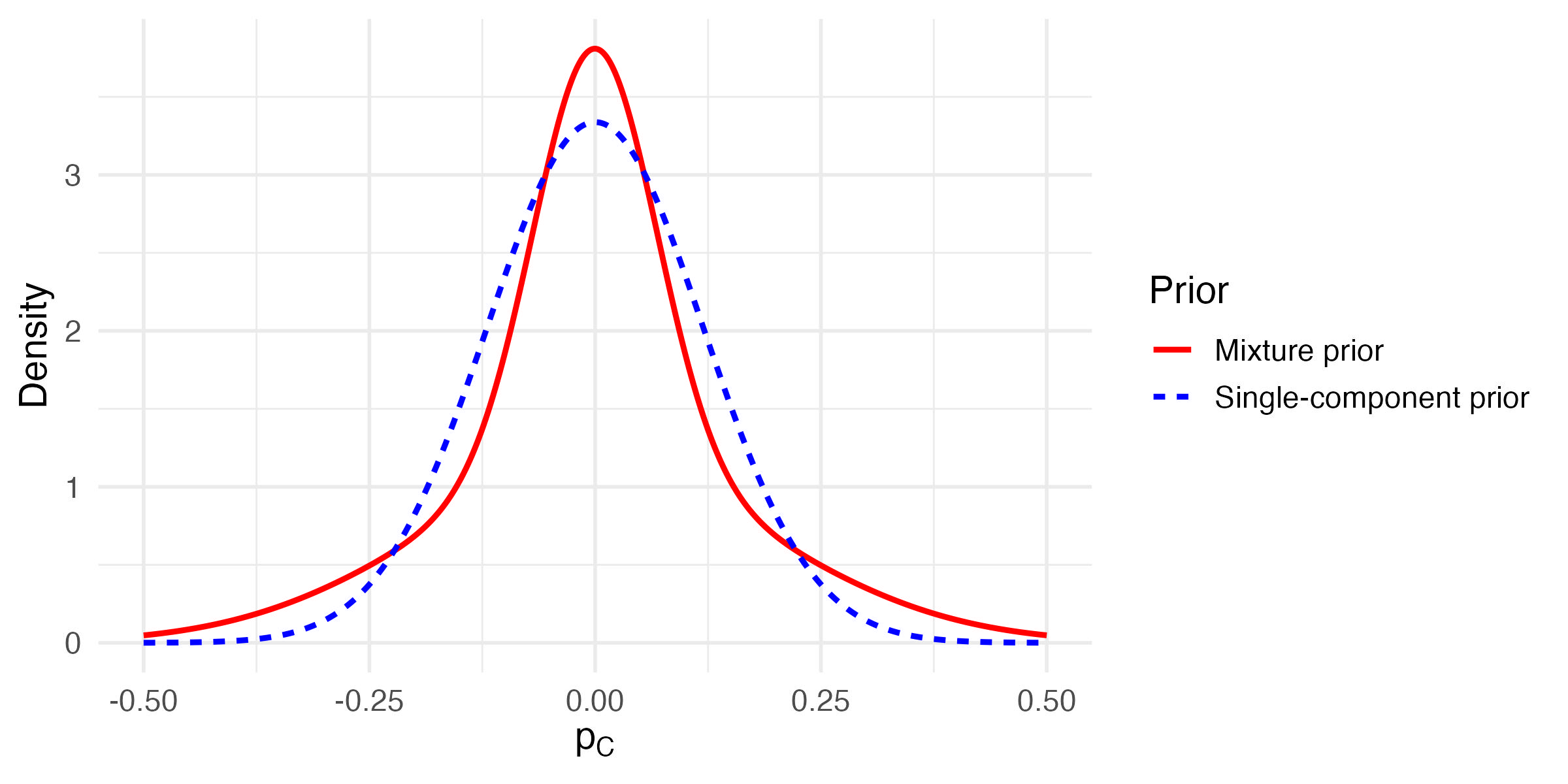}
  \caption{Prior distributions for the control parameter. The blue curve represents the single-component normal prior centered at zero with effective sample size equal to 70, which is the prior used in the main manuscript. The red curve corresponds to a mixture normal prior composed of two components with weights 0.539 and 0.461, means 0.00027 and -0.00031, and standard deviations 0.2006 and 0.0672, respectively. Both priors are defined assuming unit variance of the endpoint.}\label{fig:prior_mix_gaussian}
\end{center}
\end{figure}

We performed the same simulation study for design~1 described in Section~3.1 of the main manuscript, using a mixture prior distribution (shown in Figure~\ref{fig:prior_mix_gaussian}) centered at zero and having the same effective sample size (ESS, $n_p = 70$) as the prior used in Section~3.1.

\begin{figure}[h] 
\begin{center}
  \includegraphics[width=0.9\textwidth]{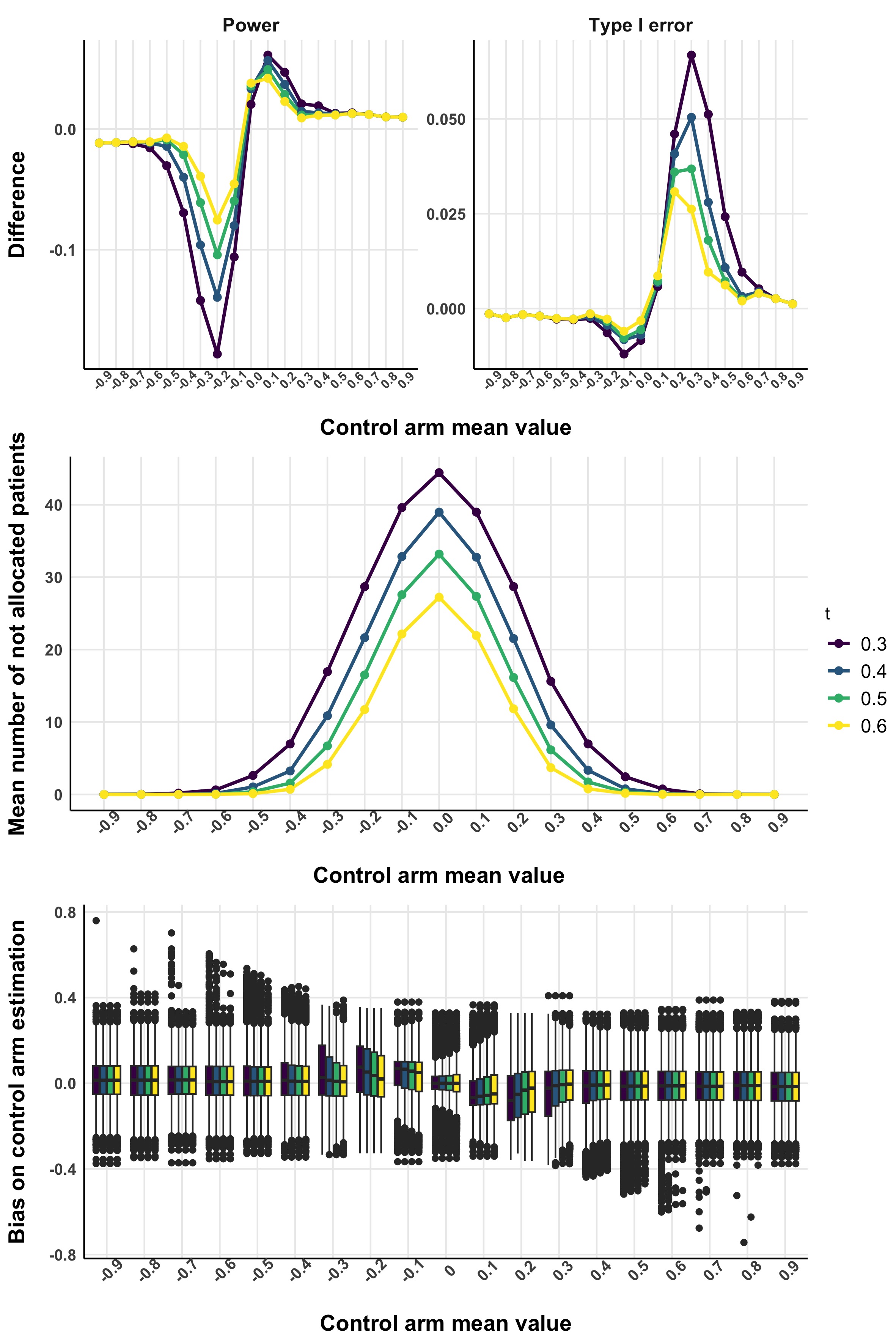}
  \caption{Results for design 1 in terms of power (top left panel), type I error (top right panel), treatment effect estimation bias (middle panel) and 95\% credible interval length (bottom panel) at different information fractions $t \in \{0.3,0.4,0.5,0.6\}$ when using a mixture prior.
  }\label{fig:resnormmix1}
\end{center}
\end{figure}

The results are shown in Figure~\ref{fig:resnormmix1}. We observe the same trend as for the non-mixture prior considered in the original manuscript.

We also investigated the performance of computing an approximate $H^*(g_C^i, \pi_C^H)$, obtained by using a Gaussian distribution with mean equal to that of the prior distribution and standard deviation equal to the unit standard deviation divided by the square root of the sample size at the interim analysis, to compute the minimal distance. Although we observed small differences in the values of $H^*$, the floor operation in Eq.~(9), that is
\[
    n^{(2)}_C = \left\lfloor\left(1-t\right)\left(1- \frac{\xi}{\lambda} \right)\frac{N}{R+1}\right\rfloor,
\]
led to identical results across all scenarios in our simulation study.

\vspace{3cm}

\clearpage

\section{Binary} 

In the binary case, patients’ responses in the treatment and control arms are modeled as Bernoulli random variables with probabilities \( p_T \) and \( p_C \), respectively. 
An initial total sample size is determined via a frequentist power calculation for a two-sample test of proportions, targeting a difference in proportions corresponding to \( p_C = 0.3 \) and \( p_T = 0.5 \), with one-sided significance level \( \alpha = 0.025 \) and power equal to 0.80.
To mimic the Gaussian case, we fix the treatment effect on the log-odds scale, even though varying \( p_C \) while maintaining the same difference in proportions would generally lead to different operating characteristics. Specifically, the treatment effect is defined as
\(
\delta = \mbox{logit}(0.5) - \mbox{logit}(0.3),
\)
and, for a given control response probability \( p_C \), the corresponding treatment response probability is defined as
\(
p_T =
\frac{\exp\bigl(\delta + \mbox{logit}(p_C)\bigr)}{1 + \exp\bigl(\delta + \mbox{logit}(p_C)\bigr)}.
\)
Under the null hypothesis, type I error is evaluated by setting \( p_T = p_C \).

An informative Beta prior distribution is specified for the control response probability, centered at \( \mu_C = 0.3 \) and corresponding to an effective sample size of 65 patients. A non-informative prior is adopted for the treatment arm, centered at \( \mu_T = 0.5 \) and equivalent to one patient. Interim analyses are conducted at information fractions
\(
t \in \{0.3, 0.4, 0.5, 0.6\},
\)
assuming equal randomization between arms in the first stage.

At the interim analysis, similarity between concurrent and historical control information is assessed using the Hellinger distance between Beta distributions. Specifically, the distance is computed between the historical control prior and the posterior distribution of the concurrent control data, after subtracting the minimal attainable Hellinger distance. The latter is obtained via numerical optimization, under the constraint that the Beta distribution has the same total information as the interim control sample. The resulting standardized distance is transformed into a borrowing weight \( 
\xi
\), with borrowing enabled only when this distance is below the prespecified threshold \( \gamma = 0.3 \). The scaling parameter \( \lambda \) is set to 1.

The amount of borrowing is translated into a reduction of the effective sample size of the control prior by rescaling both Beta parameters according to the percentage of reduction. Final inference is based on the posterior distributions of the treatment and control response probabilities, with success declared when
\[
\Pr(\Delta > 0 \mid \mbox{data}) > 0.975.
\]

\begin{figure}[h] 
\begin{center}
  \includegraphics[width=0.9\textwidth]{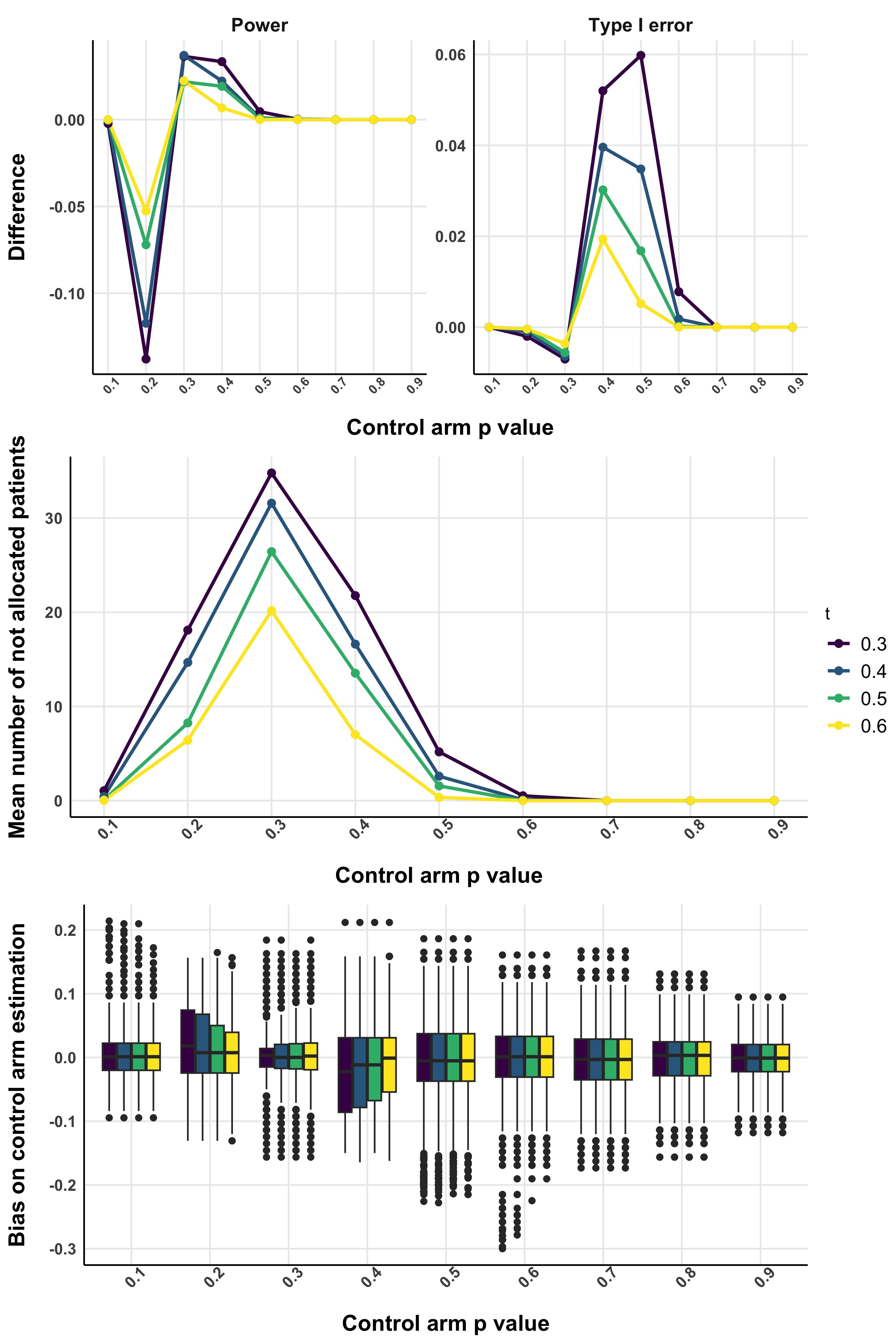}
  \caption{Results for design1 in terms of power (top left panel), type I error (top right panel), treatment effect estimation bias (middle panel) and 95\% credible interval length (bottom panel) at different information fractions $t \in \{0.3,0.4,0.5,0.6\}$.}\label{fig:resbin1}
\end{center}
\end{figure}

As a comparator, an analysis without interim analyses and without borrowing is considered, using non-informative priors for both arms (that is, using a Beta distribution centered at \( \mu_T = 0.5 \) and equivalent to one patient). A total of 5000 trials are simulated for each configuration. Performance metrics include type I error and power, posterior bias of \( \Delta \), the number of patients saved or reallocated, and the length of the 95\% equal-tailed credible interval and are shown in Figure~\ref{fig:resbin1}. All analyses are conducted in \textsf{R} (version 4.3.0), using the \texttt{RBesT} package for Bayesian modeling and effective sample size calculations.

\clearpage
\section{Meta Analytic Predictive prior for the practical application} 

\begin{figure}[h] 
\begin{center}
  \includegraphics[width=0.9\textwidth]{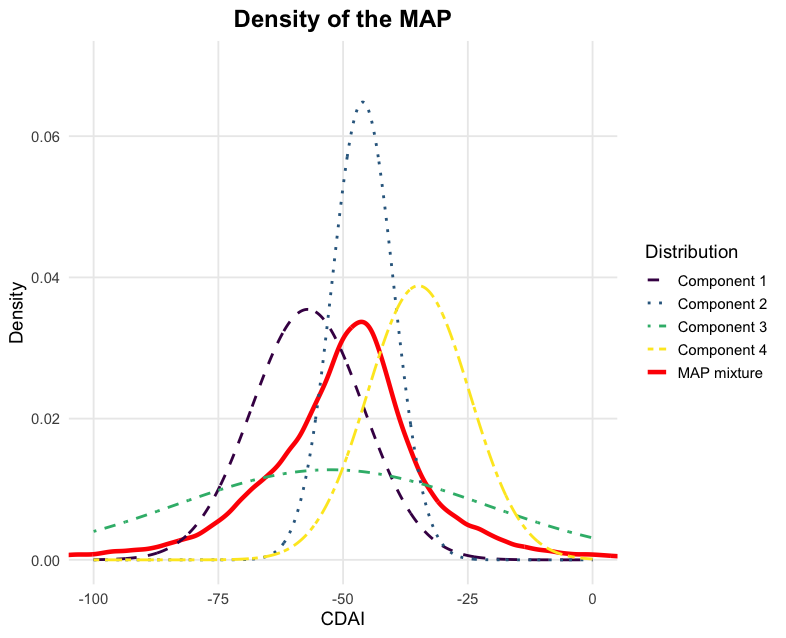}
  \caption{Density of the Meta Analytic Predictive (MAP) distribution used for the case study, approximated via a 4 components mixture distribution.}
  \label{fig:mapdensity}
\end{center}
\end{figure}

\end{document}